\input{aipcheck}

\documentclass[
    ,final            
    ,numberedheadings 
  ]
  {aipproc}

\layoutstyle{8x11double}

\usepackage{amsmath,amssymb,amsfonts}
\usepackage[utf8]{inputenc}

\newcommand{\un}[1]{\ensuremath{\ \mathrm{#1}}}
\newcommand{\uns}[1]{\ensuremath{\ \mathsf{#1}}}

\newcommand{\rsun}{\ensuremath{\ R_\odot}} 
\newcommand{\rot}[1]{\ensuremath{\nabla\times {#1}}}
\newcommand{\diver}[1]{\ensuremath{\nabla\cdot {#1}}}

\newlength{\picwd}
\begin{document}

\title{Coupling the solar surface and the corona: coronal rotation, Alfvén wave-driven polar plumes}

\classification{96.60.-j,96.60.P-,96.60.pc,96.60.Vg}
\keywords      {Sun, MHD, Solar wind, Polar plumes, Solar cycle}

\author{R. F. Pinto}{
  address={Laboratoire AIM, CEA Saclay, DSM/Irfu/SAp, U. Paris-Diderot, Gif-sur-Yvette, France},
  email={rui.pinto@cea.fr}
}

\author{R. Grappin}{
  address={LUTh, Observatoire de Paris, Meudon, France, and LPP, École Polytechnique, Palaiseau, France}
}

\author{M. Velli}{
  address={Dip. di Fisica, Unversitá di Firenze, Italy, and JPL, California Institute of Technology, Pasadena, CA, USA}
}

\author{A. Verdini}{
  address={Solar-Terrestrial Centre of Excellence-SIDC, Royal Observatory of Belgium, Brussels, Belgium}
}

\begin{abstract}
  The dynamical response of the solar corona to surface and sub-surface perturbations depends on the chromospheric stratification, and specifically on how efficiently these layers reflect or transmit incoming Alfvén waves.
  While it would be desirable to include the chromospheric layers in the numerical simulations used to study such phenomena, that is most often not feasible.
  We defined and tested a simple approximation allowing the study of coronal phenomena while taking into account a parametrised chromospheric reflectivity.
  We addressed the problems of the transmission of the surface rotation to the corona and that of the generation of polar plumes by Alfvén waves \citep{pinto_coronal_2010,pinto_coupling_2011}.
  We found that a high (yet partial) effective chromospheric reflectivity is required to properly  describe the angular momentum balance in the corona and the way the surface differential rotation is transmitted upwards.
  Alfvén wave-driven polar plumes maintain their properties for a wide range of values for the reflectivity, but they become bursty (and eventually disrupt) when the limit of total reflection is attained.

\end{abstract}

\maketitle



\section{Introduction}

The dynamics of the outer layers of the Sun strongly constraint the global topology of the solar corona and are at the origin of many transient events observed. 
The time-varying magnetic field generated by the solar dynamo connects the sub-surface layers of Sun up to the corona and heliosphere, and effectively couples the dynamics of those regions.
The photospheric movements are expected to shear the coronal magnetic arcades at their footpoints.
This shearing may be transmitted upwards by the MHD wave-modes and so affect the dynamics of the solar wind and of the corona.
The range of characteristic time and spatial scales for these surface motions is very broad.
Slow (null frequency) motions are widely believed to contribute to the quasi-steady accumulation of magnetic energy in the corona,
as required \emph{e.g.}, by the current storage-and-release type eruption models.
Finite frequency movements, on the other hand, inject waves into the same magnetic structures and provide for the vast wealth of oscillatory phenomena detected in the corona.
The standard approach to the modelling of the effects of surface movements on the corona is that of the \emph{line-tied} forcing of the magnetic loops by their footpoints, which are rigidly attached to the photosphere. 
An important consequence of this approximation is that the lower coronal boundary acts as perfect reflector to MHD waves instead of letting the chromospheric layers define the amount of reflection occurring there.
Total reflection occurs at the foot-points of the loops, which act as resonant cavities.
Finite-frequency footpoint motions induce long-lasting large amplitude oscillations \citep{berghmans_coronal_1995} and null frequency motions shear the loops by arbitrarily high amounts.
\citet{ofman_chromospheric_2002} pointed out that footpoint leakage of coronal loop oscillations could be important in some cases.
\citet{grappin_mhd_2008} have shown that the line-tying approximation is inappropriate to characterise the coronal loop's dynamical response to footpoint motions even for very high effective chromospheric reflectivities.
An alternative approximation which diametrically opposes the \emph{line-tied} condition relies on the so-called \emph{fully transparent} boundaries, as used by \citet{pinto_coronal_2010,pinto_coupling_2011}.
In these works, the authors neglected all chromospheric reflection and purposely considered ``pessimistic'' configurations in what concerns the destabilisation of coronal structures and searched for mechanisms which nonetheless are able to produce coronal events such as polar plumes, jets and inflows.
Both approximations are extreme ones, the physical reality laying necessarily in between.
The \emph{line-tied} case is prone to severely overestimate the energy input from the photosphere into the corona within magnetically connected regions, and the \emph{fully transparent} case underestimates it is the absence of the denser chromospheric (and photospheric) layers.
Ideally, one would include these layers in the numerical domain.
But that demands prohibitive increases in CPU time and model complexity, specially in multi-dimensional configurations.
Here, we test a semi-reflective lower boundary condition applied at the base of the corona which represents the effective response of the chromosphere to impinging Alfvén waves.
This condition is comparable to that used by \citet{verdini_coronal_2012} but with a different implementation, more appropriated to our MHD setup.


\section{Numerical model}
\label{sec:model}

We used the DIP numerical code, which is a 2.5 D axisymmetric model of the solar corona solving the following system of MHD equations describing a one-fluid, isothermal, fully ionised and compressible plasma:
\begin{eqnarray*}
  \label{eq:mhd}
  \partial_t \rho & + & \diver{\rho\mathbf{u}} =0\ ;\ \  P = \frac{2}{m_H}\rho k_BT \\
  \partial_t \mathbf{u} & + & \left(\mathbf{u}\cdot\nabla\right)
  \mathbf{u} = -\frac{\nabla P}{\rho} +
  \frac{\mathbf{J}\times\mathbf{B}}{\mu_0\rho} -
  \mathbf{g} + \nu\nabla^2\mathbf{u} \\ 
  \partial_t\mathbf{B} &=& \rot{\left(\mathbf{u}\times\mathbf{B}\right)} +
  \eta\nabla^2\mathbf{B}
\end{eqnarray*}
The magnetic field $\mathbf{B}$ decomposes into a time-independent
\emph{external} component $\mathbf{B^0}$ and an induced one
$\mathbf{b}$.
The solar wind develops into a stable transsonic solution
where the magnetic tension cannot counteract its dynamical pressure, hence forming the open-field regions of the corona.
The upper boundary (placed at $r=15$) is transparent.
The lower boundary (placed at $r=1.01\rsun$) is semi-reflective with respect to the Alfvén mode, but transparent with respect to all others (\emph{cf.} Sect. \ref{sec:semi}).
We recur to the description of the system of equations above in terms of the MHD characteristics at the domain's boundaries in order to achieve total transparency or to control explicitly its reflectivity.
Alfvén waves are injected at the lower boundary by perturbing the corresponding characteristic there.
The diffusive terms are adapted so that grid scale ($\Delta l$)
fluctuations are correctly damped. The kinematic viscosity is defined as
$\nu=\nu_0\left(\Delta l/\Delta l_0 \right)^2$, typically with
$\nu_0=2\times 10^{14}\un{cm^2\cdot s^{-1}}$ and
$0.01\lesssim\left(\Delta l/\Delta l_0 \right)^2\lesssim10$.
The grid is non-uniform and $\Delta l$ is minimal close to the lower boundary.
The magnetic diffusivity $\eta$ is scaled similarly.
Both coefficients were kept as low as possible in all the runs performed here (yet inevitably larger than those in the real Sun).
\citet{grappin_alfven_2000} give a more thorough description of the spatial and temporal schemes used.

\subsection{Semi-reflective lower boundary}
\label{sec:semi}

As stated in Sect. \ref{sec:model}, our lower and upper boundary conditions consist of imposing the amplitudes of the MHD characteristics propagating into the numerical domain (the outgoing ones being already completely determined by the dynamics of the system).
We treat the chromosphere as a discontinuity, hence the effective reflectivity (of the Alfvén mode) is completely determined by the ratio $\epsilon$ of Alfvén wave speeds above and below.
The chromospheric reflectivity a is then defined as
\begin{equation}
  \label{eq:a}
  a = \frac{\epsilon - 1}{\epsilon + 1}\ ,\ \mathrm{with}\ 
  \epsilon = \frac{C_A^{photosph}}{C_A^{corona}}\ ,
\end{equation}
$C_A$ representing the Alfvén speed $B/\left(\mu_0 \rho\right)^{1/2}$.
The approximation is valid for perturbations whose characteristic wavelength is much larger than the thickness of the chromosphere.
We defined two types of semi-reflective conditions, one for the Alfvén characteristic and another one for purely azimuthal motions (note that, on one hand, perturbations with purely azimuthal velocity are a mixture of MHD wave modes, and on the other hand that pure Alfvén modes are circularly polarised).
The first kind translates simply in
\begin{equation}
  \label{eq:bc_alfven}
  L_A^+ = -a\ L_a^- + \left(1+a\right) \delta L_A^+\left(t\right)\ ,
\end{equation}
where, $L_A^+$ is the amplitude of the upward-propagating Alfvén characteristic (\emph{i.e}, propagating into the domain) and $L_A^-$ is the amplitude of its downward-propagating counterpart.
The second kind assumes the following three conditions
\begin{equation}
  \label{eq:bc_vz}
  \partial_t u_\phi^+ = f\left(\theta,t\right);\ \ \partial_t u_r^+ = 0;\ \ \partial_t u_\theta^+ = 0\ ,
\end{equation}
where $f\left(\theta,t\right)$ describes a perturbation exerted at the surface.
This results in a combination of perturbations of the slow-mode, the fast-mode and Alfvén mode characteristics, all defined as a function of $f\left(\theta,t\right)$.
The two kinds of semi-reflective boundaries are used here with different purposes.
We use the first kind in Sect. \ref{sec:plumes}, which addresses a problem where the injection and reflection of pure and finite-frequency Alfvén waves is of importance.
We favour the second kind of semi-reflective boundary in Sect. \ref{sec:rotation}, on which we study quasi-steady solar wind solutions for which a better control of the surface azimuthal velocity is required (and the effects of perturbing non-Alfvénic modes are not important).

\section{Coronal rotation}
\label{sec:rotation}

\begin{figure*}

  \includegraphics[ width=0.60\picwd]{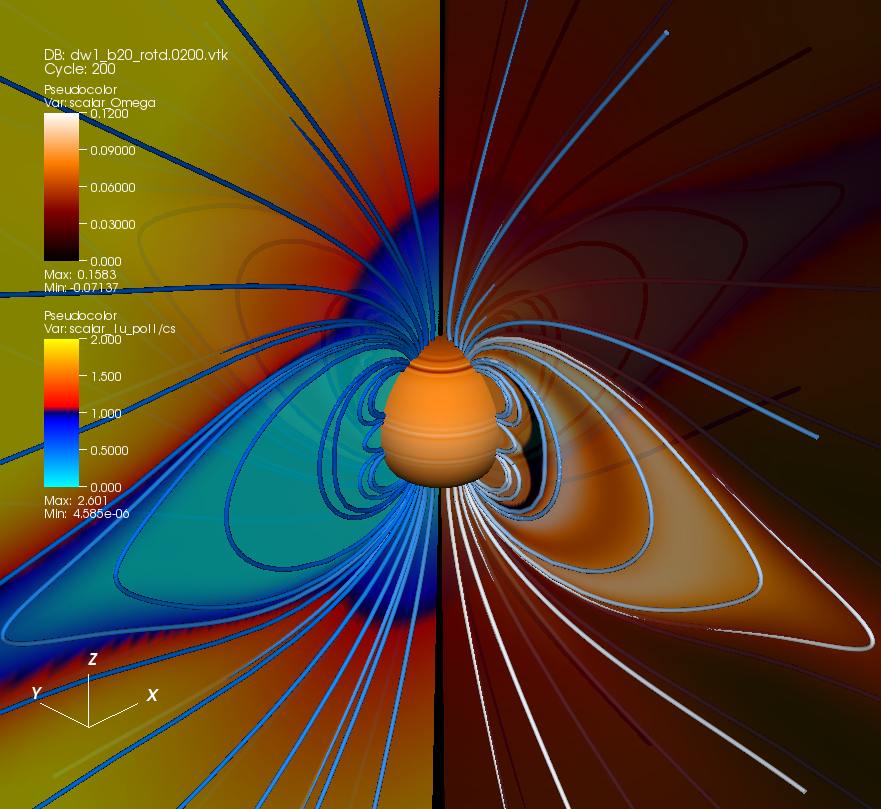}
  \includegraphics[height=0.55\picwd]{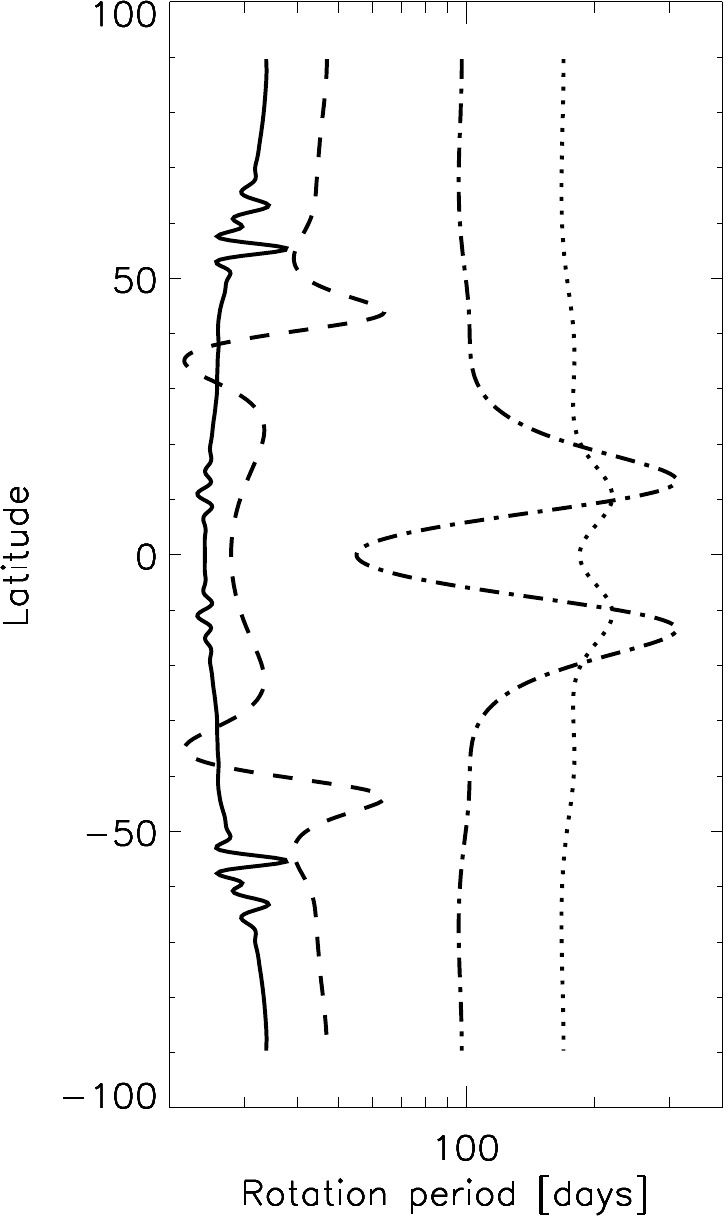}
  \hspace{0.04\textwidth}
  \includegraphics[ width=0.60\picwd]{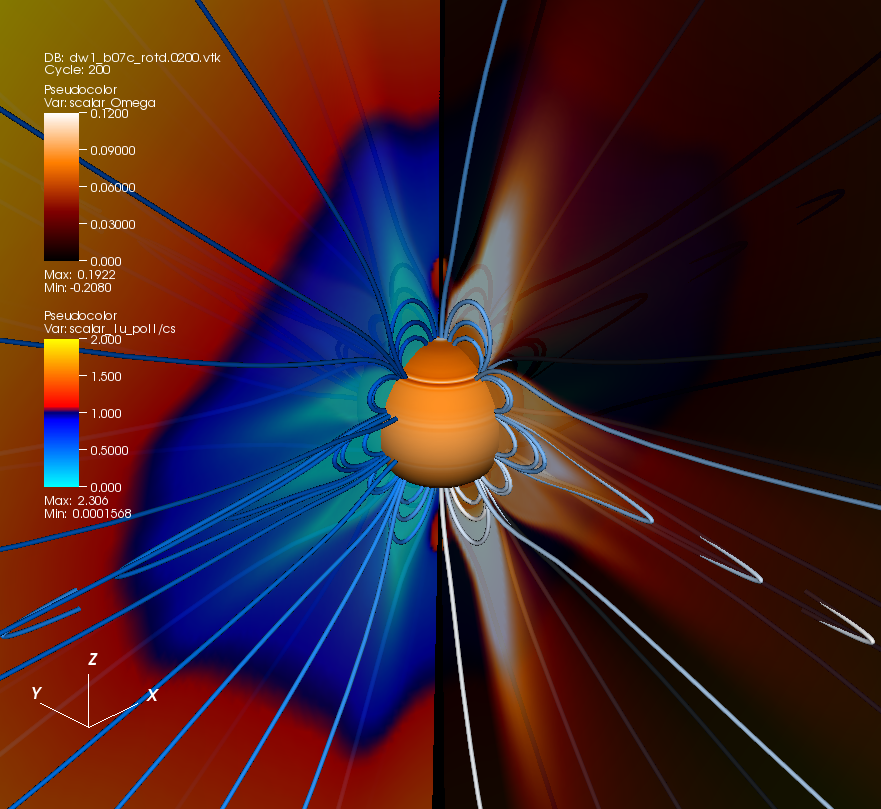}
  \includegraphics[height=0.55\picwd]{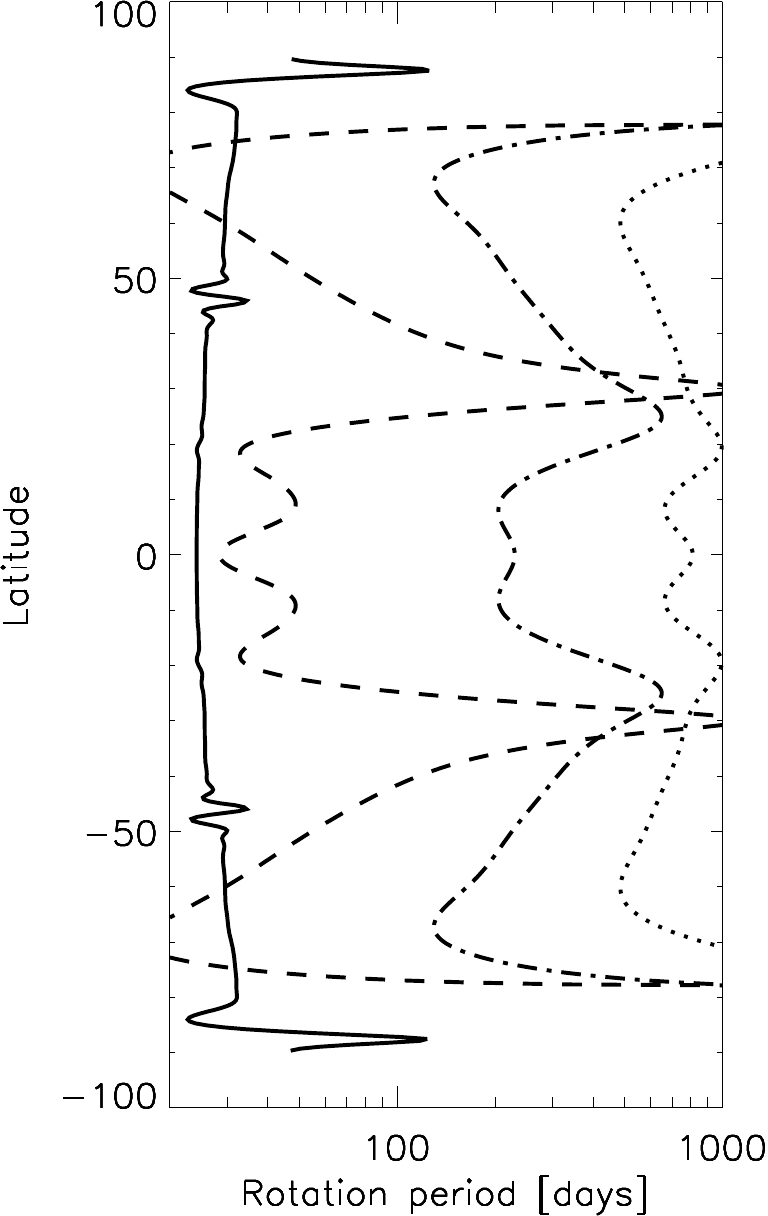}
  \caption{Coronal magnetic field, wind velocity and rotation rate at two illustrative moments of the solar cycle ($t=0$ and $t=3.8\un{yr}$ in Fig. 3 of \citep{pinto_coupling_2011}).
    The blue lines are magnetic field-lines.
    The blue-yellow colour-scale represents the wind speed in Mach number (blue: $v_r/c_s < 1$; yellow: $v_r/c_s > 1$).
    The orange colour-scale represents the rotation rate $\Omega$.
    The plots on the right side represent the rotation period as a function of latitude at different radii ($r = 1,\ 2.5,\ 7,\ 15\rsun$).
  }
  \label{fig:rotation}
\end{figure*}

\citet{pinto_coupling_2011} studied the response of the solar wind and corona to the slow (quasi-steady) variations of the global magnetic field during a solar cycle.
They produced a 11 yr time-series of magnetic fields generated by a 2.5D axisymmetric kinematic dynamo code \citep[STELEM,][]{jouve_role_2007} which was used to constraint the magnetic coronal topology in the coronal MHD numerical code DIP.
They quantified the temporal evolution of the poloidal distribution of the wind velocity, mass and angular momentum fluxes, which were found to depend strongly on the magnetic topology (hence on the moment of the solar cycle).
The determination of the angular momentum flux relied on the assumption of a solid-body and time-invariant solar rotation.
We revisit here this work and apply semi-reflective conditions to the lower boundary of the numerical domain.
We selected two illustrative instants of the \citet{pinto_coupling_2011} simulations.
These correspond to $t=0$ (activity minimum) and $t=3.8\un{yr}$ (activity maximum), and are represented in the first and fourth panels of their Fig. 3.
We started off from a non-rotating state and applied a torque at the surface which accelerated it to the following rotation rate profile
\begin{equation}
  \label{eq:rotation_rate}
  \Omega\left(\theta\right) = \Omega_a + \Omega_b \sin^2\theta + \Omega_c\sin^4\theta,\ 
\end{equation}
with $\theta$ being the latitude, $\Omega_a = 14.713\un{^\circ/day}$, $\Omega_b = -2.396\un{^\circ/day}$ and $\Omega_c = -1.787\un{^\circ/day}$.
Equation \eqref{eq:rotation_rate} fits the time-averaged differential rotation pattern at the surface of the Sun \citep{snodgrass_rotation_1990}.
The duration of the initial acceleration was $\sim1/4$ of the average (final) rotation period at the surface.
The initial transient propagates upwards as an Alfvén wavefront, accelerating the open field plasma in the azimuthal direction and exciting a few global oscillations in the closed field regions.
After a few Alfvén transit times, and for a small enough $\epsilon$, the corona and wind settle down into a quasi-steady state.
For $\epsilon\sim 1$, the highly transparent surface spins down quickly.
Note that the rotating wind flow carries a net outward angular momentum flux, and that a magnetic braking torque is applied at the surface in result.
For $\epsilon\sim 0$, the surface maintains the imposed rotation, smaller closed loops keep oscillating resonantly for a long time and the larger ones are sheared indefinitely.
Intermediate solar-like values allow for the surface rotation to be maintained in the open-flux regions while the minimal required amount of footpoint leakage is allowed for the closed-flux regions to stabilise.
The streamers gradually assume a nearly rigid rotation, while the open field regions (coronal holes) shape up as a Parker's spiral.
The rotation rate of the former is determined by that of its magnetic footpoints.
Large streamers like that on the first panel of Fig. \ref{fig:rotation} encompass a wide latitudinal range, and therefore a wide range of footpoint rotation rates.
As a result, the surface differential rotation pattern reflects more visibly in such streamers than on the smaller ones.
On the overall, the strongest latitudinal shears appear at the streamer/coronal hole frontiers.
The finite magnetic resistivity affects the width of these regions, but has a negligible effect on the overall rotation rates.
Figure \ref{fig:rotation} represents the coronal magnetic field geometry, the wind speed (in units of Mach number) and the coronal rotation rate for $t=0$ and $t=3.8\un{yr}$.
The plots to the right of the figures shows the rotation period as a function of latitude at different radii ($r = 1,\ 2.5,\ 7$ and $15\rsun$).
We note that in order to achieve long-lasting (\emph{i.e}, stable) coronal rotation profiles we had to resort to particularly low values for $\epsilon$.
The calculations presented here were made with $\epsilon=10^{-3}$ rather than a more solar-like $\epsilon=10^{-2}$.
The reason for this is that after the initial and short-lived forcing, the surface was left to evolve under the influence of the braking torque exerted by the wind flow alone without any counter-balancing torque due to the underlying convective dynamics.
This issue will be addressed in a forthcoming paper.

\section{Polar plume generation}
\label{sec:plumes}

\begin{figure}
  \centering
  \includegraphics[width=.85\picwd,clip=true,trim=0 7 0 12]%
  {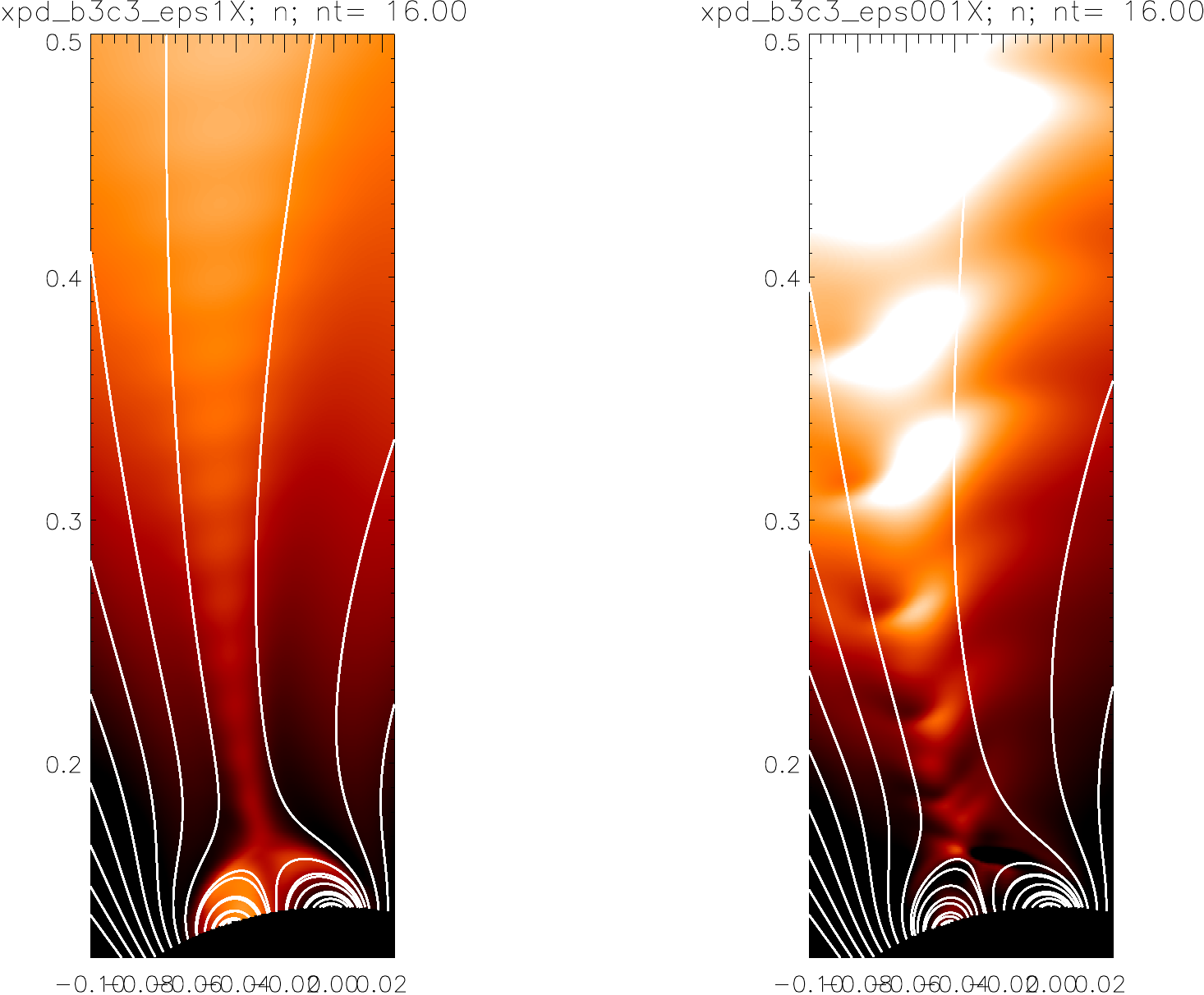}
  \caption{
    Left panel: the standard case, with $\epsilon=1$.
    Right panel: $\epsilon=0.01$.
    The colour-scale represents the density.
    White lines are magnetic field-lines.
  }
  \label{fig:plumeseps}
\end{figure}

\begin{figure}
  \centering
  \includegraphics[width=.65\picwd]%
  {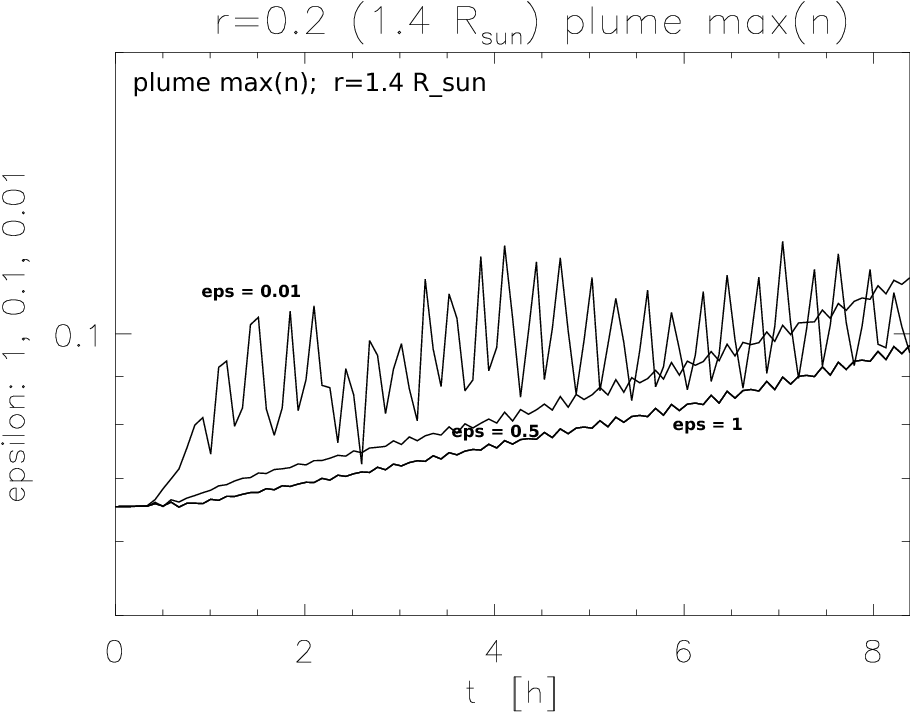}
  \caption{
    Density $n$
    as a function of time at the bottom of the jet ($r=0.2\sim
    1.4\uns{\rsun}$).
    Each curve corresponds to a given $\epsilon$.
    Note the regime change happening for $\epsilon=0.01$.
    The ordinates are in log-scale.}
  \label{fig:varepsilon}
\end{figure}

\citet{pinto_coronal_2010} studied the properties of polar plumes generated by Alfvénic torsional motions of the footpoints of a bipolar magnetic structure in a coronal hole using a fully transparent lower boundary.
Such magnetic structures appear naturally in unipolar regions such as the polar coronal holes.
They are also visible in simulations such as that in Fig. \ref{fig:rotation}.
We extended this work by accounting for the wave reflection occurring at the dense chromospheric layers by means of the reflectivity parameter $\epsilon$.
As before, Alfvén waves are continuously injected at the coronal base and drive the formation of a dense jet (plume) along the magnetic axis of the structure.
The density and velocity contrast between the axis of the plume and the background wind flow decays with height, unlike in thermally driven plume models \citep{pinto_time-dependent_2009}.
The jet displays a series of ``blobs''
(not to be confused with streamer blobs \citep{wang_origin_1998}) 
propagating outwards along its axis, which are mostly visible in the first ~7 solar radii and can be identified as slow mode wave-fronts generated non-linearly at the base of the plume \citep[\emph{cf.}][]{ofman_solar_1997,ofman_solar_1998}.
The growth rate now depends both on the frequency of the surface motions and the effective chromospheric reflectivity.
Figure \ref{fig:plumeseps} shows a snapshot of the density at $t=7.5\un{h}$ after the beginning of the wave injection for the case with $\epsilon=1$ (the standard case with transparent lower boundary) and $\epsilon=10^{-2}$.
The colour-scale represents the density corrected by a factor $\left(r/\rsun\right)^2\times \exp\left(1/\rsun-1/r\right)$ to compensate for the background stratification and let the density structures be visible.
The white-lines are magnetic field-lines.
Figure \ref{fig:varepsilon} shows the temporal evolution of the density at the axis of the plume, just above the null point.
Each curve corresponds to a different $\epsilon$.
The system's evolution remains qualitatively similar for $0.1 \leq \epsilon \leq 1$, but changes dramatically for $\epsilon \leq 0.1$.
The flow along the axis switches to a regime where it is dominated by the slow-mode wave-fronts (the ``blobs'') rather than by the smooth overdense upflow.
For $\epsilon\approx 0$, the plumes are easily disrupted.

\section{Conclusions}

We studied two different problems concerning the coupling between surface movements and the solar corona.
The first one consists of the transmission of the surface rotation to the corona and the second one of the generation of polar plumes by Alfvén waves.
For this purpose, we implemented and tested a semi-reflective lower boundary condition for our coronal model which accounts for the Alfvén wave partial reflection taking place in the chromosphere.
The effective chromospheric reflectivity is parametrised by the ratio of Alfvén speeds across the chromosphere and transition region.
We found that a high (yet partial) effective reflectivity is required for sustaining the coronal rotation against the solar wind magnetic breaking torque, while still allowing for the necessary amount of footpoint leakage.
Future work will consider the underlying time-dependent convective surface dynamics. 
We verified that the generation of polar plumes by Alfvén waves works well beyond the initially proposed limit of full transparency \citep{pinto_coronal_2010}.
The plume properties are maintained for a wide range of values for the reflectivity, but become bursty (and eventually disrupt) when the limit of total reflection is reached.
Additional work is in progress to evaluate more finely the parameter range $0<\epsilon\leq 10^{-2}$.






\bibliographystyle{aipproc}   

\bibliography{mnemonic,/data/BIBTEX/refs}

\IfFileExists{\jobname.bbl}{}
 {\typeout{}
  \typeout{******************************************}
  \typeout{** Please run "bibtex \jobname" to optain}
  \typeout{** the bibliography and then re-run LaTeX}
  \typeout{** twice to fix the references!}
  \typeout{******************************************}
  \typeout{}
 }

\end{document}